\documentstyle[preprint,aps]{revtex}

\tightenlines

\begin{document}
\title{ENERGY ASSOCIATED WITH SCHWARZCHILD BLACK HOLE IN A MAGNETIC
UNIVERSE\thanks{%
{\ This paper is dedicated to Professor George F. R. Ellis on the occasion
of his 60th birthday.}}}
\author{S.~S.~Xulu\thanks{%
E-mail: ssxulu@pan.uzulu.ac.za}}
\address{Department of Applied Mathematics, University of Zululand,\\
Private Bag X1001, 3886 Kwa-Dlangezwa, South Africa}
\maketitle

\begin{abstract}
In this paper we obtain the energy distribution associated with the Ernst
space-time (geometry describing Schwarzschild black hole in Melvin's
magnetic universe) in Einstein's prescription. The first term is the
rest-mass energy of the Schwarzschild black hole, the second term is the
classical value for the energy of the uniform magnetic field and the
remaining terms in the expression are due to the general relativistic
effect. The presence of the magnetic field is found to increase the energy
of the system.
\end{abstract}




\section{Introduction}

\label{sec:intro} Nahmad-Achar and Schutz\cite{NS} discussed that the
conserved quantities such as the energy, momentum and angular momentum play
a very important role as they provide a first integral of the equations of
motion. These help to solve difficult problems, for instance, collisions,
stability properties of physical systems etc. Obviously, it is desirable to
incorporate these quantities in general relativity. However, the study of
energy localization or quasi-localization in general relativity has been an
intractable problem and is still a subject of active interest . The energy
content in a sphere of radius $r$ in a given space-time gives a feeling of
the effective gravitational mass that a test particle situated at the same
distance from the gravitating object experiences. Recently , Virbhadra\cite
{Vir98} discussed the importance of this subject in the context of the {\em %
Seifert conjecture} as well as the well-known {\em hoop conjecture} of
Thorne. Attempts aimed at obtaining a meaningful expression for
local/quasi-local energy have given rise to different definitions of energy
and have resulted in a large number of definitions in the literature (see 
\cite{CoRi}- \cite{CN} and references therein). Canonical energy-momentum,
derived from variational formulations of general relativity, leads to
non-unique pseudotensor expressions\cite{CoRi}. Einstein's energy-momentum
complex, used for calculating the energy distribution in a general
relativistic system, was followed by many prescriptions: e.g. Landau and
Lifshitz, Papapetrou, Weinberg, and many others (see in \cite{ACV96}). Most
of these prescriptions restrict one to make calculations using ``Cartesian''
coordinates. Coordinate independent definitions of energy have been proposed
by Komar\cite{Komar}, Penrose\cite{Penrose}, and many others (see in \cite
{Brown}). Bergqvist\cite{Bergq} studied seven different definitions of
quasi-local masses for the Reissner-Nordstr\"{o}m and Kerr space-times and
came to the conclusion that no two of the definitions studied gave the same
result. In trying to bring the number of suggested quasilocal energies under
control, Hayward \cite{Hayw} listed a number of criteria which should be
followed in defining quasilocal masses. Although the subject of energy
localization has problems which still remain unresolved, some interesting
results have been found in recent years. Virbhadra and his collaborators
considered many space-times and have shown that several energy-momentum
complexes give the same acceptable result for a given space-time. Virbhadra 
\cite{KSV90} showed that for the Kerr-Newman metric several definitions
yield the same result. Following Virbhadra, Cooperstock and Richardson \cite
{CoRi} performed these investigations up to the seventh order of the
rotation parameter and reported that several definitions yield the same
result. Rosen and Virbhadra\cite{RV93} and Virbhadra\cite{Vir95} studied the
energy distribution in the Einstein-Rosen space-time and got the acceptable
result. Similarly, for several other well-known space-times it is known that
different energy-momentum complexes give the same result (see \cite{many}-
\cite{Xulu2} and references therein). Cooperstock\cite{Coop94}, Rosen\cite
{Rosen}, Cooperstock and Israelit\cite{CI95} initiated the study of the
energy of the universe. Rosen\cite{Rosen} obtained the total energy of the
closed homogeneous isotropic universe and found that to be zero, which
supports the studies by Tryon\cite{Tryon}. Aguirregabiria et al. \cite{ACV96}
showed that several energy-momentum complexes give the same result for any
Kerr-Schild class metric. The above results are favourable for the
importance of energy-momentum complexes. In this paper, we compute the
energy distribution in the Ernst space-time.

Melvin's magnetic universe\cite{Melvin} is a solution of the
Einstein-Maxwell equations corresponding to a collection of parallel
magnetic lines of force held together by mutual gravitation. Thorne\cite
{Thorne} studied extensively the physical structure of the magnetic universe
and investigated its dynamical behaviour under arbitrarily large radial
perturbations. He showed that no radial perturbation can cause the magnetic
field to undergo gravitational collapse to a space-time singularity or
electromagnetic explosion to infinite dispersion. Later, Ernst\cite{Ernst}
obtained the axially symmetric exact solution to the Einstein-Maxwell
equations representing the Schwarzschild black hole immersed in the Melvin's
uniform magnetic universe. Virbhadra and Prasanna\cite{VirPra} studied the
spin dynamics of charged particles in the Ernst space-time. In this paper we
obtain the expression for energy distribution in the Ernst space-time. We
use the convention that Latin indices take values from 0 to 3 and Greek
indices values from 1 to 3, and take $c=G=1 $. 

\section{The Einstein-Maxwell equations and the Ernst solution}

\label{sec:ernst} 

The Einstein-Maxwell equations are

\begin{equation}
R_{i}^{\ k}-\frac{1}{2}\ g_{i}^{\ k}R=8\pi T_{i}^{\ k},  \label{eq1}
\end{equation}
\begin{equation}
\frac{1}{\sqrt{-g}}\left( \sqrt{-g}\ F^{ik}\right) _{,k}=4\pi J^{i}\ ,
\label{eq2}
\end{equation}
\begin{equation}
F_{ij,k}+F_{j,k,i}+F_{k,i,j}=0,  \label{eq3}
\end{equation}
where the energy-momentum tensor of the electromagnetic field is 
\begin{equation}
T_{i}^{\ k}=\frac{1}{4\pi }\left[ -F_{im}F^{km}+\frac{1}{4}\ g_{i}^{\
k}F_{mn}F^{mn}\right] .  \label{eq4}
\end{equation}
$R_{i}^{\ k}$ is the Ricci tensor and $J^{i}$ is the electric current
density vector.

Ernst\cite{Ernst} obtained an axially symmetric electrovac solution ($J^{i}=0
$) to these equations describing the Schwarzschild black hole in Melvin's
magnetic universe. The space-time is 
\begin{equation}
ds^{2}=\Lambda ^{2}\left[ (1-\frac{2M}{r})dt^{2}-(1-\frac{2M}{r}%
)^{-1}dr^{2}-r^{2}d\theta ^{2}\right] -\Lambda ^{-2}r^{2}\sin ^{2}\theta
d\phi ^{2}  \label{eq5}
\end{equation}
and the Cartan components of the magnetic field are 
\begin{eqnarray}
H_{r} &=&\Lambda ^{-2}B_{o}\cos \theta ,  \nonumber \\
H_{\theta } &=&-\Lambda ^{-2}B_{o}\left( 1-2M/r\right) ^{1/2}\sin \theta ,
\label{eq6}
\end{eqnarray}
where 
\begin{equation}
\Lambda =1+\frac{1}{4}B_{o}^{2}r^{2}\sin ^{2}\theta .  \label{eq7}
\end{equation}
$M$ and $B_{o}$ are constants in the solution. 
The non-zero components of the energy-momentum tensor are 
\begin{eqnarray}
T_{1}^{\ 1} &=&-T_{2}^{\ 2}=\frac{B_{o}^{2}\left( 2M\sin ^{2}\theta -2r\sin
^{2}\theta +r\right) }{8\pi \Lambda ^{4}r},  \nonumber \\
T_{3}^{\ 3} &=&-T_{0}^{\ 0}=\frac{B_{o}^{2}\left( 2M\sin ^{2}\theta
-r\right) }{8\pi \Lambda ^{4}r},  \nonumber \\
T_{1}^{\ 2} &=&-T_{2}^{\ 1}=\frac{2B_{o}^{2}\left( 2M-r\right) \sin \theta
\cos \theta }{8\pi \Lambda ^{4}r}.  \label{eq8}
\end{eqnarray}
The Ernst solution is a black hole solution ($r=2M$ is the event horizon).
For $B_{o}=0$ it gives the Schwarzschild solution and for $M=0$ it gives the
Melvin's magnetic universe. The magnetic field has a constant value $B_{o}$
everywhere along the axis. Ernst pointed out an interesting feature of this
solution. Within the region $2m<<r<<B_{0}^{-1}$, the space is approximately
flat and the magnetic field approximately uniform, when $|B_{0}m|<<1$. 

To get meaningful results for energy distribution in the prescription of
Einstein one is compelled to use ``Cartesian'' coordinates (see \cite{ACV96} 
\cite{RV93}, \cite{Vir95}, \cite{many} and \cite{Rosen}).  It was believed
that the results are meaningful only when the space-time studied is
asymptotically Minkowskian. However, recent investigations of Rosen and
Virbhadra\cite{RV93}, Virbhadra\cite{Vir95}, and Aguirregabiria {\em et al.}
\cite{ACV96} showed that many energy-mometum complexes can give the same and
appealing results even for asymptotically non-flat space-times.
Aguirregabiria {\em et al.} showed that many energy-momentum complexes give
the same results for any Kerr-Schild class metric. There are many known
solutions of the Kerr-Schild class which are asymptotically not flat. For
example, Schwarzschild metric with cosmological constant. The general energy
expression for any Kerr-Schild class metric obtained by them immediately
gives $E = M - (\lambda/3) r^3$ where $\lambda$ is the cosmological
constant. This result is very much convincing. $\lambda > 0$ gives repulsive
effect whereas $\lambda < 0$ gives attractive effect.

The line element $(\ref{eq5})$ is easily transformed to ``Cartesian''
coordinates $t,x,y,z$ using the standard transformation 
\begin{eqnarray}
r &=& \sqrt{x^2 + y^2 + z^2},  \nonumber \\
\theta &=& \cos^{-1}\left(\frac{z}{\sqrt{x^2 + y^2 + z^2}}\right),  \nonumber
\\
\phi &=& \tan^{-1} (y/x) .  \label{eq9}
\end{eqnarray}
The line element in $t,x,y,z$ coordinates becomes 
\begin{eqnarray}
ds^{2} &=& \Lambda^2 (1-\frac{2M}{r})dt^2 - \left[ \Lambda^2 \left( \frac{%
ax^2}{r^2} \right) + \Lambda^{-2} \left( \frac{y^2}{x^2 + y^2} \right) %
\right] dx^2 - \left[ \Lambda^2 \left( \frac{ay^2}{r^2} \right) +
\Lambda^{-2} \left( \frac{x^2}{x^2 + y^2} \right) \right] dy^2  \nonumber \\
&-& \Lambda^2 \left[ 1 + \frac{2Mz^2}{r^2(r-2M)} \right] dz^2 - \left[
\Lambda^2 \left( \frac{2axy}{r^2} \right) + \Lambda^{-2} \left( - \frac{2xy}{%
x^2 + y^2} \right) \right] dxdy  \nonumber \\
&-& \Lambda^2 \left[ \frac{4Mxz}{r^2 (r-2M)} \right] dxdz - \Lambda^2 \left[ 
\frac{4Myz}{r^2 (r-2M)} \right] dydz ,  \label{eq10}
\end{eqnarray}
where 
\begin{equation}
a = \frac{2M}{r-2M} + \frac{r^2}{x^2 + y^2}.  \label{eqn10}
\end{equation}

\section{The Einstein energy-momentum complex}

Einstein obtained an energy-momentum complex $\Theta_i^k$, which satisfies
the local conservation laws (see in \cite{Moller}) 
\begin{equation}
\frac{\partial \Theta_i^k}{\partial x^k} = 0,  \label{eq11}
\end{equation}
where 
\begin{equation}
\Theta_i^k = \sqrt{-g} \left(T_i^k + \tau_i^k\right)  \label{eq12}
\end{equation}
$\Theta_i^k$ is referred to as the Einstein energy-momentum complex. $%
\tau_i^k$ is usually called {\em energy-momentum pseudotensor}. $T_i^k$ is
the energy-momentum tensor appearing in the Einstein's field equations.

Einstein found that 
\begin{equation}
\Theta _{i}^{k}=\frac{1}{16\pi }Z_{i,\ \ l}^{kl}  \label{eq13}
\end{equation}
where 
\begin{equation}
Z_{i}^{kl}=-Z_{i}^{lk}=\frac{g_{in}}{\sqrt{-g}}\left[ -g\left(
g^{kn}g^{lm}-g^{ln}g^{km}\right) \right] _{,m}  \label{eq14}
\end{equation}
The energy $E$ and the three components of momenta $P_{\alpha }$ are given
by the expression 
\begin{equation}
P_{i}=\int \int \int \Theta _{i}^{0}dx^{1}dx^{2}dx^{3}.  \label{eq15}
\end{equation}
thus the energy $E$ for a stationary metric is given by the expression 
\begin{equation}
E=\frac{1}{16\pi }\int \int \int Z_{0,\alpha }^{0\alpha }dxdydz.
\label{eq16}
\end{equation}
and after applying the Gauss theorem, one has 
\begin{equation}
E=\frac{1}{16\pi }\int \int Z_{0}^{0\alpha }\mu _{\alpha }dS.  \label{eq17}
\end{equation}
For $r=constant$ surface (given by $(\ref{eq9})$ ) one has the components of
a normal vector $\mu _{\alpha }=(x/r,y/r,z/r)$. The infinitesimal surface
element is $dS=r^{2}sin\theta d\theta d\phi $. 

\section{Calculations}

We have already discussed that to use the energy-momentum complex of
Einstein one is compelled to use ``Cartesian'' coordinates and therefore we
consider the Ernst metric in $t,x,y,z$ coordinates, expressed by the line
element $(\ref{eq10})$. The determinant of the metric tensor is given by 
\begin{equation}
g = - \Lambda^4  \label{eq18}
\end{equation}
The non-zero contravariant components of the metric tensor are 
\begin{eqnarray}
g^{00} &=& \Lambda^{-2}\frac{r}{r-2M},  \nonumber \\
g^{11} &=& \Lambda^{-2} \left[\frac{2Mx^2}{r^3} - \frac{x^2}{x^2+y^2}\right]
- \Lambda^2 \left[\frac{y^2}{x^2 + y^2}\right],  \nonumber \\
g^{12} &=& \Lambda^{-2} \left[\frac{2Mxy}{r^3} - \frac{xy}{x^2+y^2}\right] +
\Lambda^2 \left[\frac{xy}{x^2+y^2} \right],  \nonumber \\
g^{22} &=&-\Lambda^{-2} \left[\frac{2My^2}{r^3} - \frac{y^2}{x^2+y^2}\right]
- \Lambda^2 \left[\frac{x^2}{x^2+y^2}\right],  \nonumber \\
g^{33}&=& \Lambda^{-2} \left[\frac{2Mz^2}{r^3} - 1 \right],  \nonumber \\
g^{13}&=&\Lambda^{-2} \left[\frac{2Mxz}{r^3} \right],  \nonumber \\
g^{23}&=&\Lambda^{-2} \left[\frac{2Myz}{r^3} \right].  \label{eq19}
\end{eqnarray}

The only required components of $Z^{ij}_k$ in the calculation of energy are
the following: 
\begin{eqnarray}
Z^{01}_0 &=& \frac{4Mx}{r^3} + ( \Lambda^4 - 1) \left[ \frac{x}{x^2+y^2} %
\right],  \nonumber \\
Z^{02}_0 &=& \frac{4My}{r^3} + ( \Lambda^4 - 1) \left[ \frac{y}{x^2+y^2} %
\right],  \nonumber \\
Z^{03}_0 &=& \frac{4Mz}{r^3}.  \label{eq20}
\end{eqnarray}
Now using $(\ref{eq20})$ with $(\ref{eq17})$ we obtain the energy
distribution in the Ernst space-time. 
\begin{equation}
E = M + \ \frac{1}{16\pi} \int^{\pi}_{\theta=0}\int^{2\pi}_{\phi=0}
(\Lambda^4 - 1) r \sin\theta d\theta d\phi  \label{eq21}
\end{equation}
We substitute the value of $\Lambda$ in the above and then integrate. We get 
\begin{equation}
E = M + \frac{1}{6} B^2_or^3 + \frac{1}{20} B^4_o r^5 + \frac{1}{140} B^6_o
r^7 + \frac{1}{2520} B^8_o r^9 .  \label{eq22}
\end{equation}
The above result is expressed in geometrized units (gravitational constant $%
G =1$ and the speed of light in vacuum $c =1$). In the following we restore $%
G$ and $c$ and get 
\begin{equation}
E = M c^2 + \frac{1}{6} B^2_or^3 + \frac{1}{20} \frac{G}{c^4}B^4_o r^5 + 
\frac{1}{140} \frac{G^2}{c^8}B^6_o r^7 + \frac{1}{2520} \frac{G^3}{c^{12}}%
B^8_o r^9 .  \label{eq23}
\end{equation}
The first term $M c^2$ is the rest-mass energy of the Schwarzschild black
hole, the second term $\frac{1}{6} B^2_o r^3$ is the well-known classical
value of the energy of the magnetic field under consideration, and rest of
the terms are general relativistic corrections. For very large $B_o r$, the
general relativistic contribution dominates over the classical value for the
magnetic field energy. As mentioned in Section $2$, the gravitational field
is weak for $2m << r << B_o^{-1}$ (in $G = 1, c = 1$ units). Thus in the
weak gravitational field we have $B_0 r << 1$ ; therefore, the classical
value for the magnetic field energy will be greater than the general
relativistic correction in these cases. 

\section{ Discussion and Summary}

The energy-momentum localization subject has been associated with much
debate. Misner et al.\cite{MTW} argued that the energy is localizable only
for spherical systems. Cooperstock and Sarracino\cite{CS} contradicted their
viewpoint and argued that if the energy is localizable in spherical systems
then it is also localizable for all systems. Bondi\cite{Bondi} noted that a
nonlocalizable form of energy is not admissible in relativity. Therefore its
location can be found. Several quasi-local mass definitions were proposed
(notably by Penrose and Hawking). However, these have some problems (see\cite
{Bergq} and \cite{Vir98}). The viewpoints of Misner et al. discouraged
further study of the energy localization problem. The energy-momentum
complexes are not tensorial objects and one is compelled to use
``Cartesian'' coordinates. Due to these reasons, this subject remained in an
almost ``slumbering'' state for a long period of time and was re-opened by
remarkable results obtained by Virbhadra and Virbhadra and his collaborators
(Rosen, Parikh, Chamorro and Aguirregabiria). These works motivated many to
come back to this very interesting and important subject (for instance, see 
\cite{CoRi},\cite{CN}, \cite{Xulu1}- \cite{Rosen},\cite{BanSen}, \cite
{Yangetal} and references given there). Although the energy-momentum
complexes are not tensorial objects, they obey conservation laws (for
example, see Eq. (12)) which are true in all coordinate systems.

In the present paper we considered the Ernst space-time and calculated the
energy distribution using the Einstein energy-momentum complex. It
beautifully yields the expected result: The first term is the Schwarzschild
rest-mass energy, the second term is the classical value for energy due to
the uniform magnetic field ($E=\frac{1}{8\pi }\int \int \int B_{o}^{2}dV$,
where $dV$ is the infinitesimal volume element, yields exactly the same
value as the second term of $(\ref{eq23})$ ), and the rest of the terms are
general relativistic corrections. The general relativistic terms increase
the value of the energy. Thus, the result in this paper is against the
prevailing ``folklore'' that the energy-momentum complexes are not useful to
obtain meaningful energy distribution in a given geometry.

\acknowledgments
I am grateful to K. S. Virbhadra for guidance, George F. R. Ellis for
hospitality at the university of Cape Town, and NRF for financial support.

\end{document}